# Towards Controllable Protein Design with Conditional Transformers

Noelia Ferruz, Birte Höcker

Department of Biochemistry, University of Bayreuth, Bayreuth, Germany.

**The 21st century is presenting humankind with unprecedented environmental and medical challenges. The ability to design novel proteins tailored for specific purposes would potentially transform our ability to respond timely to these issues. Recent advances in the field of artificial intelligence are now setting the stage to make this goal achievable. Protein sequences are inherently similar to natural languages: Amino acids arrange in a multitude of combinations to form structures that carry function, the same way as letters form words and sentences that carry meaning. Therefore, it is not surprising that throughout the history of Natural Language Processing (NLP), many of its techniques have been applied to protein research problems. In the last few years, we have witnessed revolutionary breakthroughs in the field of NLP. The implementation of Transformer pre-trained models has enabled text generation with human-like capabilities, including texts with specific properties such as style or subject. Motivated by its considerable success in NLP tasks, we expect dedicated Transformers to dominate custom protein sequence generation in the near future. Finetuning pre-trained models on protein families will enable the extension of their repertoires with novel sequences that could be highly divergent but still potentially functional. The combination of control tags such as cellular compartment or function will further enable the controllable design of novel protein functions. Moreover, recent model interpretability methods will allow us to open the 'black box' and thus enhance our understanding of folding principles. While early initiatives show the enormous potential of generative language models to design functional sequences, the field is still in its infancy. We believe that using generative text models to create novel proteins is a promising and largely unexplored field and discuss its foreseeable impact on protein design.**

## Introduction

Proteins are the universal building blocks of life playing a vital role in essentially every cellular process. The custom design of specific, efficient, and tailored proteins in a fast and cost-effective manner would have the potential to tackle many of the challenges that humankind faces today and in the future. For example, we would be able to design enzymes that metabolize plastic waste or hydrolyze polluting toxins, or create new vaccines in a timely fashion in the event of a pandemic. But despite great advances, contemporary research is still far from designing proteins as proficient as those generated naturally[1].

Protein design seeks to create custom *structures* that perform a desired function. This enormous challenge has often been referred to as the inverse protein folding problem: Instead of finding the structure that a sequence folds into, the goal is to obtain an optimal sequence that adopts a certain *fold*. Mathematically, this problem is approached with optimization algorithms that search the global minimum of a sequence-structure landscape defined by an *energy function*. Despite the relative simplicity of the most widely used energy functions[2], the number of rotamers and possible combinations at each position promotes a combinatorial explosion, and understandably most protein design packages rely on heuristic algorithms. As a consequence of this complexity - and despite remarkable recent progress[3] - the design of *de novo* proteins usually takes considerable time and effort, and the overwhelming majority of functional proteins has materialized by pre-selecting naturally-occuring scaffolds and subsequently optimizing their function in iterative rounds, as opposed to concomitantly designing sequence and structure to perform a certain function[1].

While the protein design problem has been approached with physicochemical functions that study their structures, one of the most extraordinary properties of proteins is that they entirely encode their structure and function in their amino acid sequence, and they do so with extreme efficiency. The fact that sequences alone can capture proteins' properties in the absence of biophysical constraints opens an unexplored door for protein research by exploiting Natural Language Processing (NLP) methods.

## The language of proteins

Several characteristics evidence the similarities between human languages and protein sequences, where perhaps the most obvious is their hierarchical organization. Analogous to human languages, proteins are represented by a concatenation of strings: the 20 standard amino acids. Letters then assemble to form words, and amino acids combine to form secondary structural elements or conserved protein fragments[4]. Then, as words combine to form sentences that carry meaning, fragments can assemble into different protein structures that carry a function **(Fig. 1a)**.

The origin and evolution of languages and proteins also show many parallels. Languages grow and continuously adapt, with words emerging that better reflect our evolving society. Today there are over 8,000 languages divided into more than 140 linguistic families, which originated from a common ancestral language spoken in central Africa 50,000 – 70,000 years ago[5]. Likewise, all organisms living on Earth have a (last universal) common ancestor: LUCA, a microbe that lived 4 billion years ago,[6] which already contained most modern protein domains that have developed through evolution. Inspired by linguistic approaches to reconstruct ancient vocabularies by comparing modern languages, Alva et al. identified a set of primordial peptides that trace to pre-LUCA times. These peptides have been reused across the protein sequence space in very different protein contexts, comparable to how certain root words are the ancestors of today's modern languages[7].

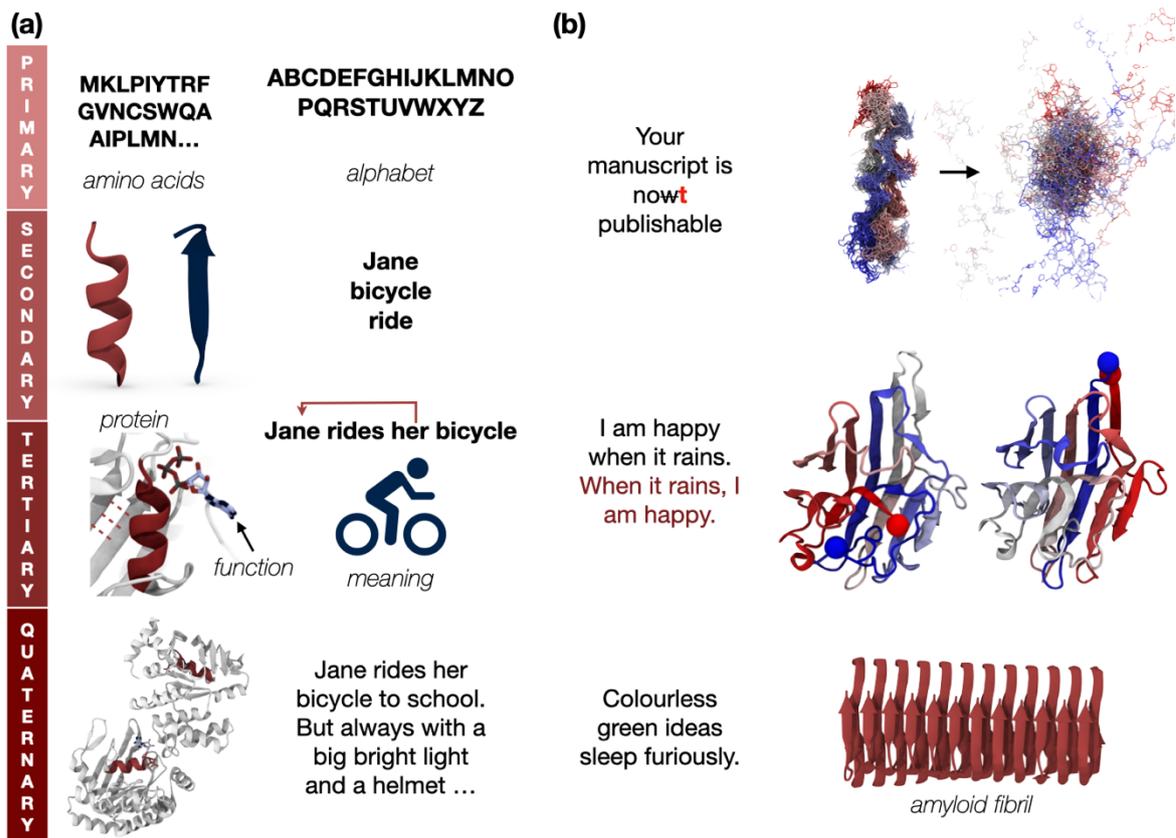

*Figure 1: Similarities between proteins and languages. (a)* Protein sequences (primary structure) are represented by a concatenation of characters of their alphabet: The 20 standard amino acids. These amino acids form three-dimensional secondary structural elements such as $\alpha$-helices and $\beta$-sheets, which like words assembling to form sentences that carry meaning, arrange to form tertiary structures that carry function. Protein domains further assemble to larger quaternary complexes similar to sentences building text. *(b)* The similarities between languages and proteins span other examples. Typos in sentences can be fatal, like missense mutations for protein functionalities. Sentences and sequences can be permuted, retaining their meaning and function, and grammatically correct sentences do not ensure a logical meaning like folded structures do not guarantee functionality.

In human languages, words bear relations and interact with adjacent words in the same way amino acids depend on their sequential surroundings. However, human languages also present long-distance dependencies, that is, any dependency between not strictly linearly adjacent words or morphemes such as subjects across sentences in long texts. This notion is reminiscent of protein structures, where amino acids far apart in the sequence could be interacting in the three-dimensional structure, sometimes crossing the domain boundaries. The associations also span other behaviours observed in proteins **(Fig. 1b).** The detrimental effect of adding or changing one letter in a sentence's meaning is equivalent to a loss of function caused by a single mutation[8]. The possibility of shuffling words while still preserving meaning is comparable to sequence permutation[9]. Lastly, the formulation of a grammatically correct but meaningless sentence is analogous to designed protein structures with no apparent function or even dangerous functions such as amyloid fibrils[10].

However, it is essential to note that protein and human languages also present dissimilarities that challenge applying NLP to protein research. We will mention here three prominent examples. Firstly, many human languages offer a clear discernible definition of words in written texts (with one prominent exception being Chinese), but the boundaries are less discernible in proteins. One possibility could be to use the boundaries of secondary structural elements (**Fig. 1a**) or conserved fragments[4]. In either case, the tokenization process would rely on the availability of tertiary structures and computationally more intensive calculations than word tokenization. A second impactful difference to modern languages is the current lack of understanding of the protein language, similar to our current lack of knowledge of many extinct languages. While we have the *Corpora* to train the protein language - unlike the case for most extinct languages -, the correct interpretation of the generated sequences will remain a challenge, requiring extensive experimental tests to decipher their functionality. Lastly, protein evolution is also obviously different from the evolution of languages, subject to randomness and environmental pressure, with a grammar that unavoidably will contain many irregularities.

Overall, these dissimilarities between human languages and protein sequences pose considerable challenges to the application of NLP to protein design. Yet despite these challenges, the apparent connections between the two fields provide a fresh perspective. Considering the exceptional environmental and medical challenges of the 21$^{st}$ century that humankind is facing, we will require innovative new approaches that transcend disciplinary borders will be required to tackle them. Current design approaches have met impressive advances but cannot yet deliver solutions that keep pace with the urgency of these problems. While these approaches will arguably continue to improve, an NLP-based viewpoint creates opportunities to gain complete control over the protein design process. We summarize in the following sections how NLP research has already influenced protein science and describe the most significant development in the field, namely, the Transformer model. The sections after will elaborate on how Transformer's unique generative capabilities could transform the protein design field, including the exceptionally challenging cases of non-natural enzymatic reactions and tailored novel functions. We hope the manuscript reaches both the Artificial Intelligence and Biology fields and encourages additional collaborative efforts towards developing and accepting NLP techniques for protein design. A glossary of selected terms is provided in **Box 1**.

## NLP has had an impact on protein research for decades

We are currently witnessing a revolutionary time in the field of Natural Language Processing (NLP). Software applications such as personal assistants (e.g., Apple Siri, Amazon Alexa, Google Assistant), chatbots, and translator machines such as Google Translate are reshaping how we interact with machines and go about our daily lives. NLP research has evolved from the days when the analysis of a single sentence could take minutes to today's search engines finding millions of websites within milliseconds[11].

Although not readily evident, the field of NLP has always impacted protein research by transferring techniques that arose as solutions to NLP problems to protein sequences. **Fig. 2a** summarizes the parallels between the two fields. For decades, NLP problems were approached with shallow Machine Learning methods, such as Support Vector Machines (SVMs) or Hidden Markov Models (HMMs), applied to solve text classification and labeling problems[12]. Likewise, HMMs and SVMs have been widely used in classification and labeling problems in proteins, such as fold recognition[13], sequence classification[14], cell localization[15], and are still the state-of-the-art methods for sequence homology detection[16].

> **Box 1: Glossary of selected terms.** Terms are depicted in italic in the main text.
>
> **Autoencoding models:** A model that is trained by predicting the input after masking or corrupting some of its tokens, such as a percentage of the words in a text.
> **Autoregressive models:** A model where the current prediction depends only on past behaviour. A model that only depends on the last item of the time series is a Markov process.
> **Big Fantastic Database:** A compendium of 2.5 billion protein sequences from several databases including Uniprot/TrEMBL+Swissprot clustered using MMseq2 and available at https://bfd.mmseqs.com/.
> **Corpus (pl. Corpora):** A collection of large and usually unstructured texts used for training models. Corpus sizes are usually measured in Gigabytes or tokens.
> **Embedding:** Representation of words - or sentences - in the form of a real-valued vector that encodes the meaning of the word such that the words that are closer in the vector space are similar in meaning.
> **Energy function:** A relationship between the energy of a system as a function of the position of their atoms.
> **Mutation:** An alteration of the amino acid sequence of a protein as a result of errors during DNA replication, mitosis or meoisis, or other damages to DNA.
> **Parameters:** The internal variables of the model whose value is optimized during training. They are also termed weights.
> **Peptide:** A stretch of amino acids connected by peptide bonds.
> **Perplexity:** A way to evaluate language models based on the uncertainty of the model to predict the next word, in mathematical terms it is the exponentiated average negative log-likelihood of a sequence. Typically applies to language models.
> **Primary structure** or **sequence:** The sequence of amino acids linked together to form a polypeptide chain.
> **Protein domain:** A region of a full protein (tertiary structure) that is self-stabilizing and folds independently.
> **Quaternary structure:** The arrangement of multiple folded protein chains.
> **Rosetta Energy Score:** The energy of a biomolecule calculated with an internal energy function, or score developed in Rosetta, a software for macromolecular modelling. The energy function considers the atomic interactions of the three-dimensional structure.
> **Secondary structure:** The three dimensional form of local segments of proteins, such as alpha helices or beta sheets (**Fig. 1a**).
> **SMILE:** Abbreviation for Simplified Molecular-Input Line-Entry System, a string notation describing a chemical molecular entity.
> **Tertiary structure** or **structure:** Three dimensional shape of a protein.
> **Tokenize:** The process of breaking a text or sentence into individual linguistic units. It is usually the first step in NLP when modelling data.

Since the 2010s, however, neural networks started to produce superior results in various NLP tasks (**Fig. 2b**). The popularity of convolutional networks (CNNs) exploded amongst NLP researchers due to their success in name-entity recognition (NER), part-of-speech (POS) tagging, and semantic role labeling[17]. CNNs' applicability thus soon extended to protein research for the prediction of protein disorder[18], DNA binding sites[19], and fold classification[20]. CNNs, however, failed to model long-distance information, which is essential for global text comprehension, or in the case of proteins, what would be long-range contacts. For this reason, NLP researchers switched to Recurrent Neural Networks (RNNs)[21], in particular, Long Short-Term Memory (LSTM). RNNs presented superior capabilities in learning long-term dependencies[22] and soon were used to create language models[23]. Inspired by their success, Alley et al. utilized this architecture for a protein language model, UniRep, which predicted sequence stability with higher accuracy than previous methods[24]. Traditional LSTMs were soon superseded by attention mechanisms[23], influencing recent breakthroughs in protein research such as AlphaFold[25]. Based on the attention model, Google released the Transformer[26], improving results in most NLP tasks at a much lower computational cost. The first Transformer opened a new era in NLP, and since then, a myriad of adaptations have been implemented (**Fig. 2a**). It is worth to mention the Generative Pre-Trained Transformer[27] (GPT) and the successors GPT2[28], and GPT3[29]. These pre-trained models have shown superior performance in most NLP tasks and, for the first time, were capable of generating human-like, long, coherent articles. These recent developments in the NLP field have a great potential to be adapted to protein research. The following sections will offer insight into how pre-trained language models could transform and dominate protein design in the years to come.

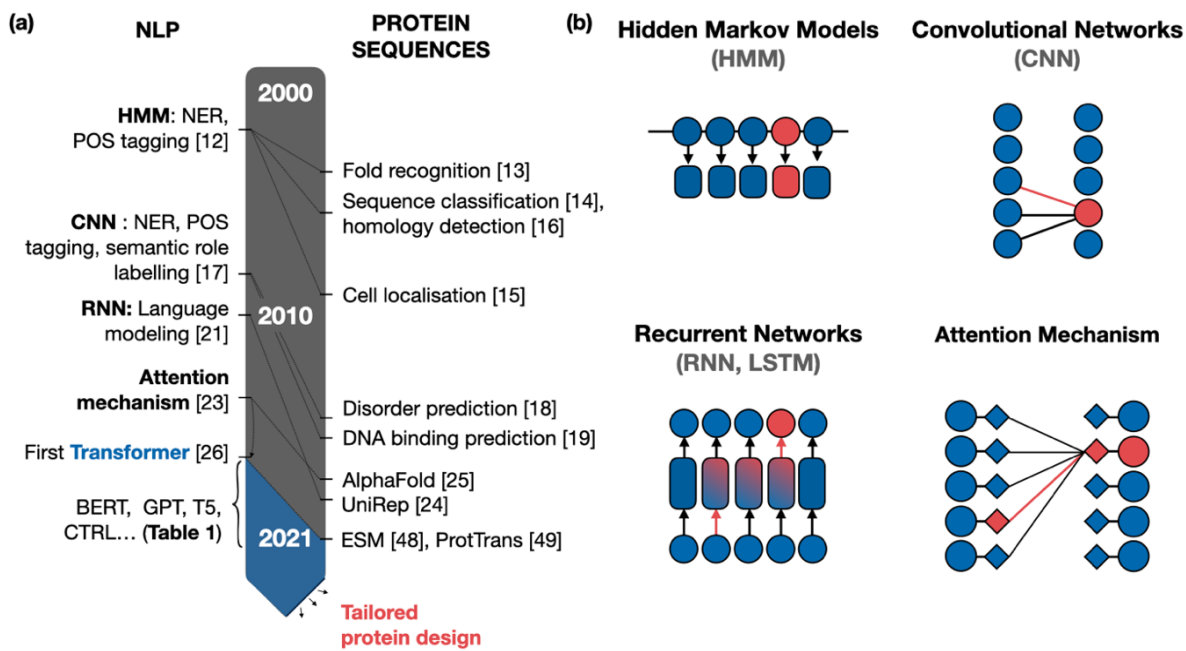

**Figure 2: Overview of most commonly used methods for NLP problems. (a)** Recent timeline of NLP methods and their application in protein research. Each breakthrough in NLP is mirrored years later in protein research applications. **(b)** Graphical explanation of the most used methods for NLP and their abbreviations. While Hidden Markov Models are stochastic processes, Convolutional Networks, Recurrent Networks, and Attention Mechanisms are or take part in neural networks.

### Attention mechanism and Transformers

Transformers are a current revolution in NLP. Their success derives from the evolution of a series of concepts built on top of each other, with the attention mechanism possibly being the most notable of these advances.

The attention mechanism originated as a solution to traditional sequence-to-sequence (seq2seq) models, widely used for tasks that process sequences from one domain to another, such as machine translation or text summarization. In seq2seq models, the input is stepwise processed in a module termed encoder to produce a context vector passed to a decoder, responsible for generating an output (**Fig. 3a**). Traditionally, encoder and decoder architectures were usually RNNs or LSTMs (**Fig. 2b**), and the context vector corresponded to the final output of the last encoder step (word) (**Fig. 2b**). Due to this inherently sequential nature, RNNs presented the major drawback of degrading performance and increasing training times with sequence length[11]. The introduction of the attention mechanism provided a solution to this problem by allowing the decoder to analyse the whole input and focus on specific parts of it, a notion similar to attention in the human mind. A simplified example in English-French translation would be focusing on the input word 'home,' when outputting the word '*maison*' (**Fig. 3a**).

While attention mechanisms had been ubiquitously applied in many types of neural networks, they became particularly prominent in 2017, when researchers at Google published their seminal work 'Attention is all you need', which introduced an architecture that not only applied attention between the modules but also throughout them[26]. This new design permitted the encoder to focus on specific parts of the input sequence, producing a much better performance in many tasks[26]. The model was termed the Transformer and gave name to all similar architectures that followed in subsequent years (**Fig. 2a**).

The Transformer's encoder and decoder modules contain a stack of six submodules or layers (N) that process inputs from the previous layer in a parallel fashion, enabling much faster training times (**Fig. 3b**). The encoder submodules contain two layers: a self-attention layer, which applies the attention mechanism to the input sequence itself, and a feed-forward layer that processes the inputs from the previous layer separately and identically, allowing parallelization. The decoder also comprises six submodules with these same two layers, but with another encoder-decoder attention layer in between that focuses on relevant parts of the input (**Fig. 3b**).

A summary of the internal workings of the dot-product attention mechanism from the original Transformer goes as follows: Because sentences are strings, the input needs to be converted to a vector of floats compatible with the internal mathematical operations of the model[30]. This step is termed *embedding*. The Transformer further adds the embeddings to positional encoding vectors that offers information about the position of each word. The final vectors are passed to the first encoder submodule, which converts them into Query, Key, and Value vectors by multiplying them with matrices obtained during training. The dot-product of the Key vector against each word's Value vector ($q_1 \cdot v_1$, $q_1 \cdot v_2$,) produces a set of scores ($s_{1,1}$) which are later scaled to the 0 to 1 range. The score multiplies the corresponding Value vectors (v1), preserving their magnitude when the score is 1 and minimizing them on the contrary. These vectors are summed up into a final output vector (z1), which represents a contribution of all the other words in the sentence for each word. In the original implementation, this process was repeated in parallel in eight attention heads (H), expanding the model's capability to focus on different input parts by creating independent Query/Key/Value vectors. A summary of the hyperparameters of this model are presented in **Table 1**.

**Table 1**: **Summary of Transformer models.** All cases report the largest Transformer of their series. Shadowed rows correspond to models that have been applied to protein sequence datasets. Data not known is depicted with a hyphen ('-'). (N = Number of transformer layers; H = number of attention heads; d= dimension of the model).

| Model | H | N | d | Training set | Parameters | Computational Time (available information) | Architecture/ training objective | Reference |
|---|---|---|---|---|---|---|---|---|
| Transformer | 8 | 6 | 512 | WMT English – German (4.5 M sentence pairs) WMT English – French (36 M sentence pairs) | - | 3.5 Days 8 NVIDIA P100 | Encoder-decoder | 26 |
| GPT | 12 | 12 | 768 | BooksCorpus (800 M words) | 110 M | 1 month 8 GPUs | Autoregressive | 27 |
| BERT (Large) | 16 | 24 | 1,024 | BooksCorpus, English Wikipedia (2,500 M words). Total: 3.3 B words or 16 GB | 340 M | 4 days 16TPU Pods (64 TPU chips) | Autoencoding | 31 |
| GPT2 | 25 | 48 | 1,600 | WebText (40 GB or 10 B Tokens) | 1558 M | 32 TPUv3 (128 chips) | Autoregressive | 28 |
| GPT3 | 96 | 96 | 12,288 | Total: 499 B Tokens from Common Crawl, WebText2, Books1, Books2, Wikipedia. | 175 B | 36 years with 8 V100 GPUs | Autoregressive | 29 |
| TransformerXL | 16 | 18 | 1,024 | Several datasets | 257 M | - | Autoregressive | 32 |
| XLNet | 16 | 24 | 1,024 | BERT dataset + 114 GB additional. Total: 130 GB. | 340 M | 512 TPU chips - 2.5 days. | Autoregressive | 33 |
| XLM | 12 | 12 | 2,048 | BERT dataset | 665 M | 64 Volta GPUs | Several learning objectives | 34 |
| RoBERTa | 16 | 24 | 1,024 | BERT dataset + 144 GB additional. Total: 160 GB | 255 M | 1024 V100 GPUs for 1 day (4-5 times more than BERT) | Autoencoding | 35 |
| DistilBERT | 6 | 12 | 768 | BERT dataset | 65 M | 8 * V100 * 3.5 days (4 times less than BERT) | Autoencoding | 36 |
| CTRL* | 16 | 48 | 1,280 | Wikipedia, Project Gutenberg, Amazon reviews, and Reddit. Total: 140 GB | 1600 M | - | Autoregressive | 37 |
| T5-11B | 128 | 24 | 1,024 | The Colossal Clean Crawled Corpus (C4, 750 GB) | 11 B | 4.68 days on 2,048 A100 GPUs (for T5-3B) | Encoder-decoder | 38 |
| Electra | 16 | 24 | 1,024 | XLnet dataset | 335M | - | Generator (autoregressive) and discriminator (Electra: predicting | 39 |

| | | | | | | | masked tokens) | |
|---|---|---|---|---|---|---|---|---|
| Albert | 64 | 12 | 4,096 | BERT dataset | 223 M | 512 TPUv3 chips 32h | Autoencoding | 40 |

Motivated by the Transformer architecture, OpenAI released GPT (Generative Pre-trained Transformer), the first of a series of highly performing pre-trained models[27]. GPT's main idea consisted of creating a general, task-agnostic language model by training it on a diverse corpus of unlabelled text followed by fine-tuning on labelled datasets to perform specific tasks, thus transferring knowledge from the first step. Despite its general nature, GPT significantly improved state-of- the-art methods in 9 out of 12 tasks studied, with the added advantage that it only required training once[27]. GPT was pre-trained on the classic language modelling task, namely, predicting the next item of a sequence based on the previous ones – a task that makes it particularly powerful for language generation. Models trained on this objective are termed *autoregressive*, and in the case of Transformers, usually, their architecture corresponds to the stack of layers from the decoder module (**Fig. 3c**). However, GPT's generative capabilities did not become evident until the implementation of GPT-2, a model with ten times more parameters and training data[28] (**Table 1**). GPT2 showed such an incredible performance at generating coherent text that the authors decided to withhold the model due to the risk of misuse, a decision that was met with controversy[41]. More recently, openAI publicized its third generation GPT model, GPT3, containing 100 times more parameters than GPT-2 (**Table 1**) and capable of performing well in a *zero-shot* fashion even on tasks it had never been trained on, like code writing[29].

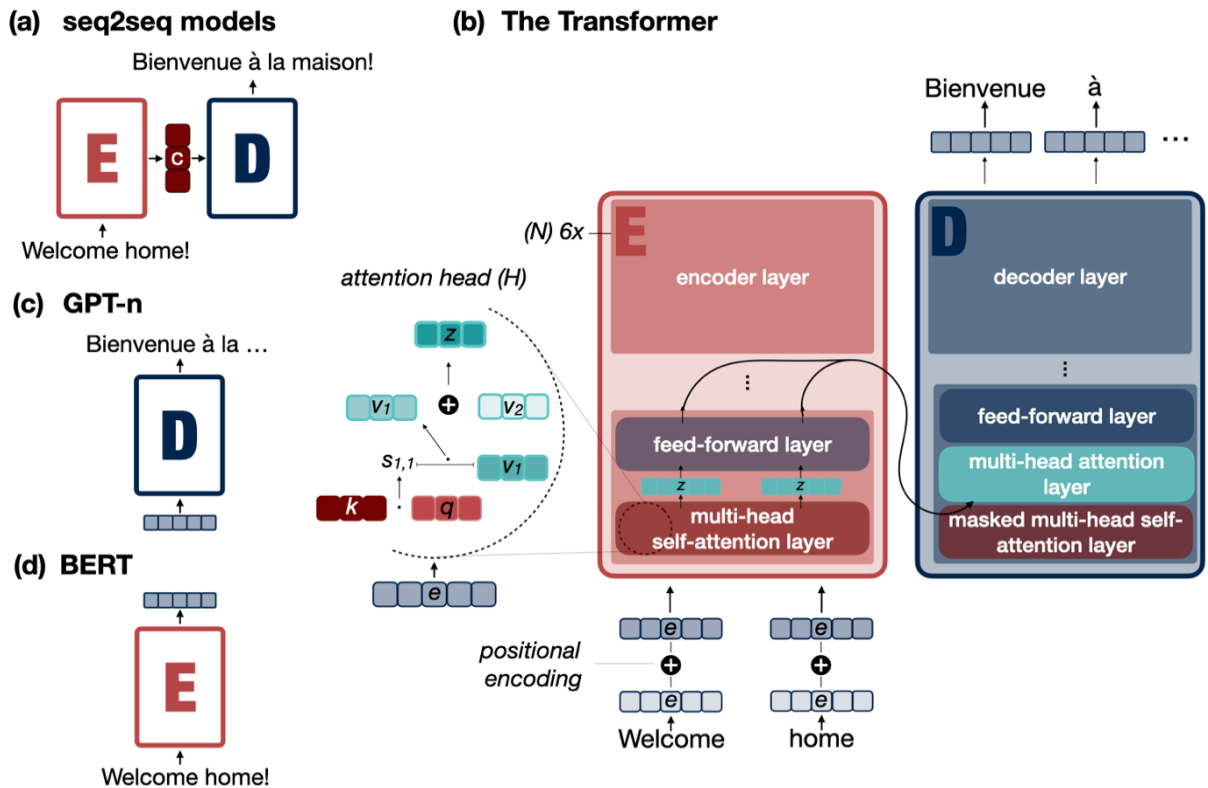

**Figure 3**: **Schematic overview of most used Transformers. (a)** seq2seq models present encoder (E) and decoder (D) models processing sequential inputs that are encoded as context (c) vectors. **(b)** The original Transformer architecture consisted of encoder and decoder models with stacks of 6 layers each. An overview of the attention dot mechanism that takes place in the attention head is presented (see main text). **(c)** The GPT-n Transformers are based on the original Transformer but contain only the decoder model, whereas BERT uses only the encoder **(d)**.

Another prominent development for the NLP field came also from the Google AI Language team, who pre-trained BERT (Bidirectional Encoder Representations from Transformers) to create a language model[31]. BERT is also inspired by the Transformer architecture but given that in this case the interest lies in creating representations of text input it only uses the encoder module (**Fig. 3d**). Models like BERT are called *autoencoders*, which are pre-trained by corrupting the input tokens in some way and trying to reconstruct the original sentence. Although they can also be used for text generation[42], they are most often applied to produce vector representations that can be used for downstream tasks such as classification[43].

In addition to these two representative examples of the encoder and decoder-only architectures of Transformers, many other Transformers have been published in the last three years. An interesting overview of these is provided by Qiu et al.[44] and a summary of most prominent Transformers' properties is provided in **Table 1**.

## Protein sequences are ideal candidates for Transformers

The amount of digital data generated is growing exponentially. In 2020 alone, we accumulated over 40 zettabytes (ZB) of data, and current estimates set the levels to more than 80 ZB by 2025[45]. Indeed, the enormous success of the last-generation Transformers comes in part to the ever-growing corpora they are trained on (**Table 1**), which, in turn, permits creating larger and more powerful models (**Fig. 4a**).

This data explosion is, however, not specific to web data. The size of biological databases is also growing exponentially, a trend most noticeable for protein sequences. **Fig. 4b** illustrates data acquisition trends in the last 20 years for sequence and structural databases, revealing that the characterization of protein sequences is growing at a much faster rate than their counterpart structures. The UniParc database, a comprehensive and non-redundant dataset containing most of the publicly available protein sequences, consists of 441,169,278 entries (release 2021_03). Although the recent development of high-performing methods for structure prediction like Alphafold[25,46] has enabled scientists to equate the growth of structures with sequences, it does not solve the time-consuming problem of functional annotation. We are thus dealing with a field where the ratio of unlabelled-to-labeled data increases exponentially (a phenomenon termed the sequence-structure gap), reminiscent of the exponential accumulation of unlabelled corpora on the internet. Given the success of semi-supervised methods such as Transformers harnessing scrapped web data to create language models, we can speculate that Transformers could similarly exploit the vast protein space and stimulate a similar revolution in the protein research field.

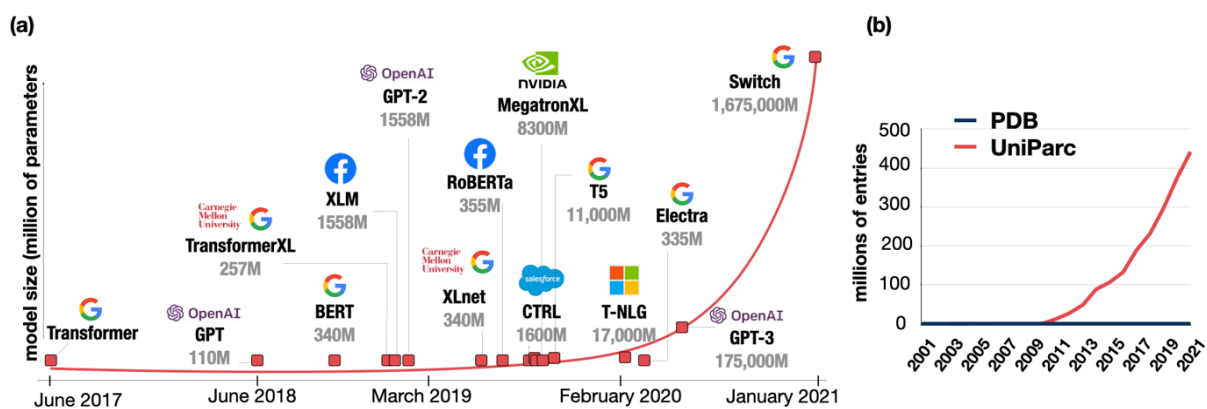

**Figure 4: (a)** Overview of most prominent Transformers released in the last years and their number of parameters. **(b)** Deposited entries in the protein databases PDB and UniParc.

## Transformers for protein design

The recent revolutionary developments in NLP are already influencing some pioneering protein research. Inspired by advances from the pre-Transformer era, several studies applied the concept of language models to protein sequences. Yu et al. applied *n-gram modelling* to generate a probabilistic protein language model[47], while in the same year, Alley et al. utilized LSTMs to implement another protein language model, UniRep, able to output vector representations of protein sequences[24]. In a similar fashion to BERT, UniRep's vector representations were used in downstream tasks, in this case allowing to predict protein sequence stability or function among others with state-of-the-art accuracy[24].

It was however not until 2021 that the first Transformer pre-trained on protein data was released. In an effort led by Facebook AI, ESM-1b, an encoder Transformer trained on 250 million protein sequences, was published[48]. ESM-1b has the same architecture and training objective as BERT (**Fig. 3d**), but in this case 33 encoder layers were pre-trained on the UniParc database (**Fig. 3d**) to produce vector representations that encode protein sequences. ESM-1b's representations, analogous to BERT sentence representations capturing language grammar, encode the internal organization of proteins from the level of biochemical properties of amino acids to that of evolutionary relationships among proteins. Coupling to downstream deep learning models enabled up-to-date predictions of mutational effects and improved predictions for long-range contacts[48].

More recently, a collaboration of scientists from Munich, Nvidia and Google AI led to ProtTrans, an impressive adaptation of six previously released Transformer-based architectures (Transformer-XL, BERT, Albert, XLnet, T5, and Electra, **Table 1**) to the protein domain and completely accessible to the community (https://github.com/agemagician/ProtTrans). This study utilized the largest training dataset to date, containing over 390 billion amino acids taken from UniParc and the *Big Fantastic Database*. Their work showed that protein embeddings – the vector representations that the Transformers output - are capable of accurately predicting per-residue secondary structure and sub-cellular localization[49].

While these early studies demonstrated the potential of learned protein representations for downstream applications including classification or regression tasks, to date there are no publicly available pre-trained models for protein sequence generation. Intentions to generate sequences from Transformers have however been recently reported. Want et al. theorized that Transformers like BERT are Markov random field language models, and as such have the potential to generate text following a Gibbs sampling procedure[42]. Following this idea, Johnson et al. implemented the protein Gibbs sampler, which coupled the previous methodology to the ESM-1b Transformer[50]. While this approach enables to directly generate sequences from ESM-1b, it is worth mentioning that Gibbs sampling from BERT produced sentences of inferior quality when compared to the decoder-only Transformer GPT[42]. Taking into account that the newest GPT-2 and GPT-3 provide better generative capabilities - in the case of GPT3 these are sentences that are often indistinguishable from human-generated ones, as recently reported in an article published in the Guardian by GPT-3 itself[51] - it could be extrapolated that GPT-2 and GPT-3-like Transformers would produce protein sequences of superior quality.

GPT's power not only resides in its generative capabilities. GPT was also finetuned and coupled to a variety of downstream tasks[27]. The inputs corresponded of sentence-label pairs, which were passed through pre-trained GPT to obtain the last layer's activations. After finetuning on this task, the representations can be fed to a linear layer to predict the labels. This is analogous to obtaining protein representations through ESM-1b which can be coupled to other models for predictions of protein properties such as stability changes or function loss. In this manner, a GPT-like architecture trained on protein sequences would provide enormous advantages for protein design, being specifically trained to generate new sequences, while retaining the possibility to obtain representations which can be coupled to other models to, for example, predict the effect of certain mutations (**Fig. 5c**).

Moreover, GPT models can be finetuned for conditional text generation. One issue with standard language generation via generative Transformers is that the direction the text takes is unpredictable. Fine-tuning Transformers on specific types of data, such as scientific publications or article news, has shown to provide a control on the type of output[52]. A decoder-only Transformer trained on protein data would offer the same capabilities, i.e, generating sequences from a particular family or fold after fine-tuning on the natural set of this group, thus extending protein families repertoires with foldable sequences of predicted similar activities. Thus, the recent advances in NLP and the development of more powerful Transformers are setting the stage to revolutionize protein research and protein design in the near future.

**Tailored protein design**

The next important step in NLP and its application to custom protein design is the recent development of the Conditional TRansformer Language (CTRL), an autoregressive model including conditional tags capable of controllably generating text[37] (**Table 1**). These tags, called control codes, allow users to more specifically influence genre, topic, or style, among others – an enormous step towards goal-oriented text generation. Shortly after CTRL implementation, the authors adapted this model to a dataset of 281 million protein sequences[53]. The model, named ProGen, contains as conditional tags UniparKB Keywords, a vocabulary of 10 categories including 'biological process,' 'cellular component,' or 'molecular function.' In total, the conditional tags comprised more than 1100 terms. ProGen presented *perplexities* representative of high-quality English language models, even on protein families not present in the training set. Generation of random sequences and their Rosetta energy evaluation revealed that the sequences had better scores than random ones. The authors analyzed ProGen's capabilities to complete a truncated kinase domain and showed that all completed proteins remained close to the Rosetta score of the native protein. As the last test for generative capabilities, several protein G-binding domain variants passed through ProGen – selection of the top one hundred variants with the lowest perplexity values provided better fitness scores than random mutations. In a later application of this work, the authors applied ProGen to the generation of lysozymes after fine-tuning on five different protein families. Experimental validation showed that the generated sequences possess enzymatic activities in the range of natural lysozymes, while X-ray characterization of one of the variants showed that it recapitulated the native three-dimensional structure[54].

These three studies highlight a promising new area of research: the controlled generation of protein sequences with conditional Transformers. The inclusion of conditional tags in Transformer-based protein language models will not only enable the generation of novel sequences as in these previous works but could potentially also provide control over the properties of these proteins. We will mention here a few possibilities.

Firstly, we envision to directly generate sequences that have a property included in the training set, such as binding ATP, folding into an all-beta structure, or being membrane-bound (**Fig. 5d**). Secondly, it would be important to investigate property tags that appear in several regions across protein space, such as 'membrane protein' or 'ATP binding' (**Fig. 5e**). The output sequences would render so far unknown solutions for these properties - proteins in unexplored regions of the sequence space (**Fig. 5e**) - and provide the means to understand their structural requisites for these functional characteristics. And lastly, controlled Transformers would enable the tailored design of proteins with novel functions. Analogous to how the combination of control tags, such as style + topic ('poetry' + 'politics'), provides specific text, the fusion of protein properties could create novel functions, such as 'hydrolase + 'PET binding' or 'membrane bound' + 'protease.'

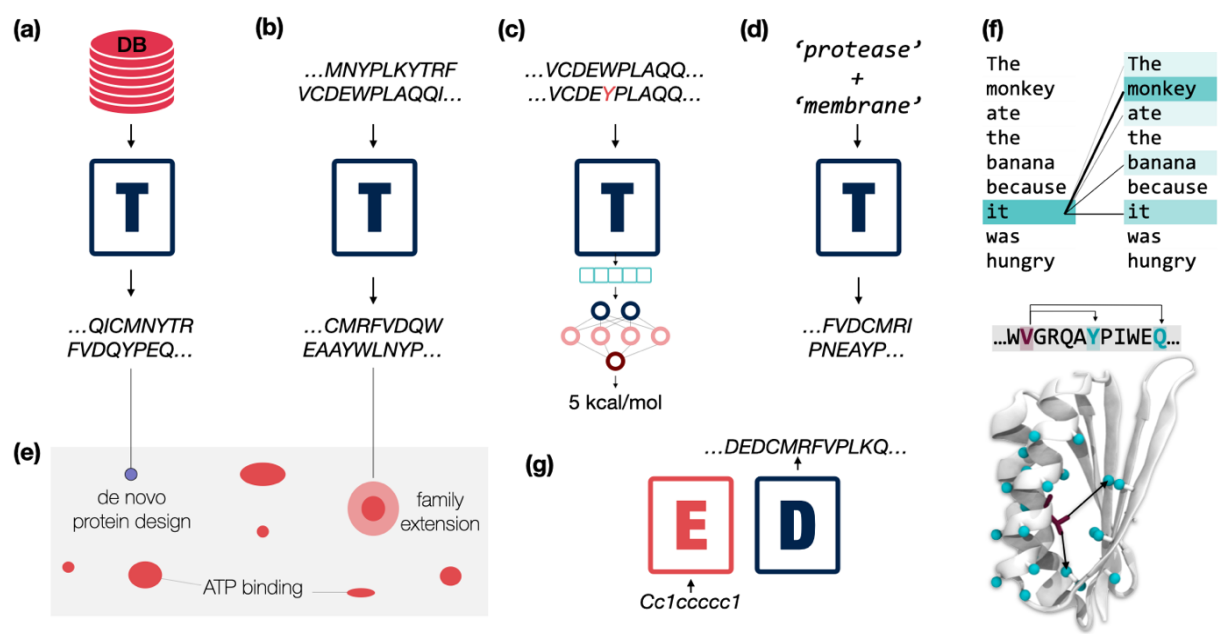

**Figure 5: Overview of possibilities in the protein engineering field with Transformer models. (a)** After training a Transformer (T) with a protein sequence database, it is possible to generate *de-novo* protein sequences **(e)**. **(b)** Fine-tuning the pre-trained model on a protein family would generate novel sequences compatible with the family. **(c)** Last layer's vector representations can be used in a variety of downstream tasks by training them with coupled models, to for example, predict protein stability. **(d)** Controlled Transformers will be capable of generating sequences with certain properties, such as 'protease' or 'membrane'-bound. **(e)** A schematic view of protein space, adapted from [55]. **(f)** Visualization of the attention mechanism has opened the door to understand Transformer models, which along with other techniques, could be used to understand protein design principles, such as required interactions. **(g)** Machine translation models such as from the original transformer could enable receptor and enzyme design.

**Enzyme, receptor, and biosensor design**

In 2018, IBM Research released IBM RXN for Chemistry, a cloud-based app that relates organic chemistry to human language[56]. The app hosts the Molecular Transformer, a model that can predict the most likely outcome of a chemical reaction using an encoder-decoder architecture (**Fig. 3a**). In this case, the model encoder processes chemical species as input (reactants + reagents), whereas the decoder outputs the most likely reaction products[56]. Subsequently, the authors reversed the network: Instead of predicting the outcome of a possible chemical reaction, the problem consisted of determining the reactants needed to create a given target molecule, a process termed retrosynthesis[57]. Following a similar approach, Grechishnikova implemented an encoder-decoder architecture for *de novo* drug generation[58]. In this case, the encoder processes protein sequences while the decoder generates *SMILES* of ligands that are potentially compatible with binding the input sequence.

These two examples show how models based on the original Transformer are powerful tools for generating outputs conditioned on an entry input. In particular, Grechishnikova's approach is interesting for the protein design realm, whereby reversing the translation machine we might be capable of generating sequences compatible with the encoder input SMILES (**Fig. 5f**). Such a model could have tremendous applications for engineering of receptor proteins, including the prediction of sequences for the recognition and binding of specific ligands, a big step forward for receptor and biosensor design. Given the recent approach by IBM to encode vector representations of chemical reactions[59], we could envision another model that takes chemical reactions as input and produces protein sequences as output. Such a model would provide an innovative route for enzyme design, including engineering enzymes capable of catalyzing reactions not found in nature. This approach could potentially support biological strategies e.g., to reverse environmental pollution.

**Explainable protein design**

The design of proteins with customizable properties is a long-standing goal in Biochemistry. On a more fundamental level, there is also the interest in understanding the principles that relate sequences to protein structures, which would enable the rational design of funnel-shaped protein folding energy landscapes[60]. For this reason, there is a growing interest in providing interpretations for the underlying mathematical workings of deep learning models in a way that is understandable to the human mind. Explainable artificial intelligence (XAI) would help us understand why models reach a particular answer and lead scientists to new ideas and approaches. Research in the drug discovery field is already benefitting from the application of XAI techniques, for example to identify ligand pharmacophores that drive a molecule's activity[61].

Traditionally, the most widely used NLP techniques, such as HMMs or SVMs (**Fig. 2**) were inherently explainable and are therefore attributed with the term 'white box'. The recent explosion of deep learning methods reaching high performance across NLP tasks has brought the challenge of developing new techniques to explain these models. Significant progress has been made on XAI techniques for 'black box' models, among which the five main techniques are feature importance, surrogate model, example-driven, provenance-based, and declarative induction. For a recent review covering these techniques in the NLP domain, we refer to Danilevsky et al[62].

For the particular case of Transformers, the use of attention mechanism throughout their architectures provides advantages for explaining their internal representations. The attention mechanism itself corresponds to an importance score over the input features, which allows visualizing the raw scores as a saliency heatmap. **Fig. 5f** exemplifies the self-attention for a sentence where a particular attention layer has attributed several attention scores between the word "it" and others. In an analogous fashion, protein sequences would correspond to a representation of attention scores among the amino acids (**Fig. 5f**). Recently, efforts have been made to bring XAI for Transformers into user-friendly interfaces. For example, exBERT (https://exbert.net)[63] enables visualization of internal representations for any Transformer trained on any corpus. It is possible to visualize self-attention user-defined sentences for all the different attention layers, select specific words and visualize the network Part-of-speech prediction at each layer, or search them over the training corpus showing the highest-similarity matches. An adaptation of exBERT to a protein-trained Transformer would enable interactive visualization of relationships among amino acids in a protein and, similar to POS-tags, their predicted properties. Similarly, searching protein fragments in the training corpus and finding the highest-similarity matches could illuminate new relationships between proteins. Although this field is still in its infancy, the possibility of visualizing the internal workings of Transformers could bring great opportunities to better understand protein folding and design.

**Is the future of protein design in the hand of big companies?**

The landscape of Transformer models published in the last years is dominated by big companies (**Fig. 4a**). Training GPT-3 with 175 B parameters -the second-largest model to date- was estimated to have cost $12 million and required over 10,000 days of GPU time[64]. Other models have been trained by accessing large *TPU* resources. Training such deep learning models is a commodity that might be accessible to large companies such as OpenAI or Google, but is potentially beyond the reach of start-ups and academic research groups. Not only their economic accessibility is a concern, but also the carbon footprint associated with training such AI models is getting growing attention[65]. Although there is increasing awareness of these possible problems associated with AI, the truth is that models perform considerably better with increasing size[66], and model sizes will most predictably only

continue to grow: The amount of computing used in the largest AI training runs has been increasing exponentially this year at a 3.4-month doubling time rate[67].

This has obvious repercussions for protein research and academic groups. The 7 protein-based Transformer models published to-date (**Table 1**) all correspond to efforts led by or including large companies. While this might sound like a troubling prospect for academic groups and the overall future of this rapidly evolving field, this does not necessarily create an imbalance.

First, large Transformer models have the advantage of only requiring training once and can then be used for a wide variety of downstream tasks, suggesting that the research community would still benefit after public release. Examples of this are efforts including AlphaFold and ProtTrans, but unfortunately public release is not always the case. Moreover, while the protein-based published Transformers correspond to big companies' efforts, in all cases they involved collaborations with academic groups, a trend that if extended in the future might bring new opportunities to academia and create a more collaborative research community, with science and ultimately society benefitting from the funding opportunities brought on by large companies. Lastly, while large language models tend to perform better, there have been also efforts to implement equally performing models with lesser computational resources, such as DistilBERT, which retains 97% of BERT performance while reducing its size by 40% and Switch, which has up to 7x increase in pre-training speed with the same computational resource than T5. These last examples are reminiscent of the analysis of long-timescale molecular dynamics, which initially was only accessible to companies with costly, specialized-hardware like ANTON[68], but soon became accessible to the whole research community with the use of in-house GPU clusters and elegant algorithmic solutions[69,70].

## Conclusion

The recent developments in the NLP field and its potential applications to protein sequences are opening exciting new doors for protein research and the design of customizable proteins. Transformer-based language models have served a variety of tasks, including translating natural language, or even writing code to train machine learning models. Moreover, these new models have been capable of generating text so similar to humans that since their inception they have been surrounded by controversy, often not being released due to concerns about potential misuse in the form of fake news or unethical medical advice[41]. Regardless of these discourses, these examples clearly show the incredible potential of Transformers. Given the similarities between language and protein sequences, the protein research field will undoubtedly benefit from this transformational new technology[71].

We envision six direct applications from transferring current NLP methods to the protein research domain, as summarized in the previous sections and illustrated in **Fig. 5.** Ordered by how readily applicable current NLP Transformers are to protein sequences, we could (1) generate sequences in unobserved regions of protein space, (2) fine-tune sequences of natural protein families to extend their repertoires, (3) utilize their encoded vector representations as input for other downstream models for protein engineering tasks, (4) generate conditional sequences with specific functional properties, (5) design completely novel and purpose-driven receptors and enzymes using encoder-decoder Transformers and (6) gain a more complete understanding of sequence-structure-function relationships and the rules that govern protein folding by interpreting these language models. Given these promising opportunities, we believe that Transformer-based protein language models will revolutionize the field of protein design and provide novel solutions for many current and future societal challenges. We hope that our ideas reach both the Artificial Intelligence and Biochemistry communities and encourage application of NLP methods to protein research.

## Competing interests
The authors declare no competing interests.